\documentstyle[aps,preprint]{revtex}
\newcommand{\select}[2]{#1}

\begin{document}

\draft
\tighten

\preprint{LANCS-TH/0001} %% hep-th/0001209
\title{Cosmology of Randall-Sundrum models with an extra dimension
stabilized by balancing bulk matter}
\author{Hang Bae Kim}
\address{Department of Physics, Lancaster University, Lancaster LA1 4YB, UK\\
{\tt h.kim@lancaster.ac.uk}}
\maketitle
\begin{abstract}
We provide the cosmological solutions for Randall-Sundrum models
with the bulk energy-momentum $\hat T^5_5$ incorporated.
It alters the Friedmann equation for the brane scale factor.
We make a specific choice of $\hat T^5_5$
which is adjusted to stabilize the extra dimension.
This makes it possible to compactify the extra dimension with a single
positive tension brane, and this model provides a RS-type solution to the
cosmological constant problem.
When the same idea is applied to the RS model with two branes,
the wrong sign of Friedmann equation for the negative tension brane
can be resolved and
usual FRW cosmology is reproduced for the brane.
\end{abstract}

\select{}{\begin{multicols}{2}}

\section{Introduction}

There has been enormous interest in the brane world theories recently.
In these theories, it is assumed that standard model particles are confined
to a (3+1)-dimensional brane of higher dimensional spacetime, while gravity
propagates in the whole bulk.
One virtue of these models is that very large extra dimensions are allowed
without conflicting with the current experimental bounds,
which makes it possible to lower the fundamental scale of gravity
to the electroweak scale by introducing large extra dimensions
\cite{ADD}.
There has been an extensive study on phenomenology and cosmology of this
model. %\cite{}.
Randall and Sundrum (RS) proposed different brane models \cite{RS1,RS2},
in which the background metric is curved along the extra dimension
due to the negative bulk cosmological constant.
These RS models have drawn much attention recently
because they provide a new way of addressing the gauge hierarchy problem
and the cosmological constant problem.
There have been studies on
the characteristics of these models \cite{characteristic},
the extensions to higher dimensions and more branes \cite{extension}
and connections to supergravity and string theories \cite{string},
their phenomenological consequences \cite{phenomenology},
and inflationary solutions \cite{inflation,KK}
and cosmology \cite{BDL,CGKT,CGS,KKOP1,BDEL,CGRT,KKOP2,cosmology}
of these models.

The cosmology of these models can be very different
from the conventional four dimensional cosmology.
Hence it can provide a lot of interesting constraints on these models.
In regard of this,
an important observation is that, without the bulk cosmological constant and
the brane tension, the late evolution of the brane scale factor deviates from
the usual FRW universe \cite{BDL}. 
For the RS model with a positive tension brane,
it was found that the usual FRW universe can be reproduced \cite{CGKT,CGS}.
However, for the negative tension brane which is considered to be our universe
in the RS model which solves the gauge hierarchy problem, the opposite sign
appears in front of the brane energy density.
Furthermore, to maintain the extra dimension to be static in this model,
a very specific relation needs to be imposed between brane energy densities
of positive and negative tension branes as well as brane tensions themselves.
Recently, it was pointed out that this is because the stabilization mechanism
is not included in the model and these problems disappear if it is included
\cite{CGRT}.

In this paper, we attempt a simple extension of previous analyses
for brane cosmology in five dimension,
by incorporating the bulk energy-momentum,
specifically non-trivial $\hat T^5_5$ component.
It alters the evolution of the brane scale factor
and leads to intriguing consequences.
We make a very specific choice of $\hat T^5_5$
which is adjusted to stabilize the extra dimension.
Though it requires a very specific form of $\hat T^5_5$
which is correlated with brane energy densities,
there was recently an argument that it may
arise naturally as a consequence of stabilization mechanism \cite{KKOP2}.
In addition, this is a minimal extension in which the extra dimension
can be stabilized without assuming correlation between energy densities
on the branes, and the exact bulk solution can still be obtained.

We apply this idea to the RS models with one and two branes.
The existence of balancing bulk energy momentum $\hat T^5_5$
makes it possible to compactify
an extra dimension just with a single brane.
This single brane model has an interesting feature that
it converts a cosmological constant problem to a dynamical
problem to determine the size of extra dimension,
as the original RS model did for the gauge hierarchy problem.
When applied to the Randall-Sundrum's two-brane model,
the negative matter density on the negative tension brane is not required any
more because the extra dimension is now stabilized by balancing $\hat T^5_5$
for any values of energy densities.
However, if we assume the parameter values which fit the current cosmological
constant bound and the gauge hierarchy problem, we obtain a very unusual
Friedmann equation for the scale factor of negative tension brane,
whose evolution is governed mainly by the energy density of positive tension
brane.  To reproduce the conventional Friedmann equation in this model,
the energy densities on two branes are required to be highly correlated.
This looks very striking, but contriving at first sight,
but we have no reason to discard this possibility since we started with
two brane tensions correlated in such a way in this model.
We will also mention loopholes in the model which make it free from the above
conclusion.

This paper is organized as follows:
In section \ref{section2}, we describe our framework of five-dimensional
brane models and Einstein equations.
In section \ref{section3}, we present the exact bulk solution and the Friedmann
equation for the brane scale factor in the presence of $\hat T^5_5$.
In section \ref{section4}, we explore one and two brane models,
with $\hat T^5_5$ which is tuned to stabilize the extra dimension.
We conclude and present discussion in section \ref{section5}.

\section{The Framework}
\label{section2}

We consider the five-dimensional spacetime with coordinates $(\tau,x^i,y)$
where $\tau$ and $x^i$ denote the usual four-dimensional spacetime and
$y$ is the coordinate of the fifth dimension.
The action describing our framework is
\begin{equation}
S = \int d^5x \sqrt{-{\hat g}}
\left[ \frac{M^3}{2}{\hat R} + \hat{\cal L}_M \right],
\end{equation}
where $M$ is the fundamental five dimensional Planck mass
and $\hat{\cal L}_M$ represents all non-gravitational contributions,
including those responsible for branes.
The five dimensional Einstein equation derived from this action is
\begin{equation}
\hat G_{MN}=\frac{1}{M^3}\hat T_{MN}.
\end{equation}
Since we are interested in the cosmological solutions,
we consider the metric
\begin{equation}
ds^2 = -n^2(\tau,y)d\tau^2 + a^2(\tau,y)\gamma_{ij}dx^idx^j + b^2(\tau,y)dy^2
\end{equation}
where $\gamma_{ij}$ is a 3 dimensional homogeneous and isotropic metric and
we will use $K=-1,0,+1$ to represent its spatial curvature.
The five-dimensional Einstein tensor $\hat G_{MN}$ for this metric is given by
\begin{eqnarray}
\hat G_{00} &=& 3\left\{
 \frac{\dot a}{a}\left(\frac{\dot a}{a}+\frac{\dot b}{b}\right)
-\frac{n^2}{b^2}\left[\frac{a''}{a}
    +\frac{a'}{a}\left(\frac{a'}{a}-\frac{b'}{b}\right)\right]
+ K\frac{n^2}{a^2}\right\}, \\
\hat G_{ij} &=& -\frac{a^2}{n^2}\gamma_{ij} \left\{
 2\frac{\ddot a}{a}
+\frac{\dot a}{a}\left(\frac{\dot a}{a}-2\frac{\dot n}{n}\right)
+\frac{\ddot b}{b}
+\frac{\dot b}{b}\left(2\frac{\dot a}{a}-\frac{\dot n}{n}\right)\right\}
-K\gamma_{ij} \nonumber\\
&& +\frac{a^2}{b^2}\gamma_{ij} \left\{
 2\frac{a''}{a}+\frac{n''}{n}
+\frac{a'}{a}\left(\frac{a'}{a}+2\frac{n'}{n}\right)
-\frac{b'}{b}\left(2\frac{a'}{a}+\frac{n'}{n}\right)\right\}, \\
\hat G_{55} &=& 3\left\{
-\frac{b^2}{n^2} \left[
\frac{\dot a}{a}\left(\frac{\dot a}{a}-\frac{\dot n}{n}\right)
+\frac{\ddot a}{a}\right] +
\frac{a'}{a}\left(\frac{a'}{a}+\frac{n'}{n}\right) \right\}, \\
\hat G_{05} &=& 3\left(\frac{n'}{n}\frac{\dot a}{a} + \frac{a'}{a}\frac{\dot
b}{b} - \frac{\dot a'}{a}\right),
\end{eqnarray}
where dots and primes denote derivatives with respect to $\tau$ and $y$
respectively.
The energy-momentum tensor is a function of $\tau$ and $y$, and divided into
bulk and brane sources
\begin{equation}
\hat T^M_N = {\rm diag}[-\hat\rho,\hat p,\hat p,\hat p,\hat p_5]
+ \sum_{i={\rm branes}}\frac{\delta(y_i)}{b}{\rm diag}[-\rho_i,p_i,p_i,p_i,0].
\end{equation}
Singular brane sources are treated by imposing the junction conditions
on the bulk solutions for the bulk sources
\begin{equation}
\label{BC}
\left.\frac{1}{b}\frac{n'}{n}\right|_{y_i-}^{y_i+}=\frac{2\rho_i+3p_i}{3M^3},
\quad
\left.\frac{1}{b}\frac{a'}{a}\right|_{y_i-}^{y_i+}=-\frac{\rho_i}{3M^3}.
\end{equation}

One of our assumptions about the energy-momentum tensor is that
$\hat T_{05}=0$, which means that there is no flow of matter along the fifth
dimension. This implies
\begin{equation}
\label{05}
\frac{n'}{n}\frac{\dot a}{a}+\frac{a'}{a}\frac{\dot b}{b}-\frac{\dot a'}{a}=0.
\end{equation}
With this equation, we can write the 00 and 55 equations in simpler form.
Let us define \cite{BDEL}
\begin{equation}
F(\tau,y) \equiv \frac{(a'a)^2}{b^2} - \frac{(\dot aa)^2}{n^2} - Ka^2.
\end{equation}
Then the 00 and 55 component equations are written as
\begin{eqnarray}
F' &=& -\frac{(a^4)'}{6M^3}\hat\rho, \label{F1} \\
\dot F &=& \frac{(a^4)\dot{\vphantom a}}{6M^3}\hat p_5, \label{F2}
\end{eqnarray}
If $\hat\rho$ is independent of $y$, which is the case we will consider
in this paper, we can integrate (\ref{F1}) with respect to $y$ and obtain
\begin{equation}
\label{F}
F = -\frac{a^4}{6M^3}\hat\rho + C(\tau),
\end{equation}
where $C(\tau)$ is an integration constant which does not depend on $y$.
Differentiating (\ref{F}) with respect to $\tau$ and using (\ref{F2}),
we get the equation for $C$
\begin{equation}
\label{C}
\dot C = \frac{(a^4)\dot{\vphantom a}}{6M^3}(\hat p_5+\hat\rho)
+\frac{a^4}{6M^3}\dot{\hat\rho}.
\end{equation}

Before we try to solve bulk equations, let us mention the energy-momentum
conservation equation, $\partial_M\hat T^M_N=0$.
\begin{eqnarray}
& \displaystyle
\label{EMC1}
\frac{d\hat\rho}{d\tau} + 3(\hat\rho+\hat p)\frac{\dot a}{a}
+(\hat\rho+\hat p_5)\frac{\dot b}{b} = 0,
&\\& \displaystyle
\label{EMC2}
\hat p_5'+\hat p_5\left(\frac{n'}{n}+3\frac{a'}{a}\right)
+\frac{n'}{n}\hat\rho - 3\frac{a'}{a}\hat\rho = 0.
&
\end{eqnarray}
These equations will give constraints on $\hat p_5$.

\section{Cosmological solutions with a static extra dimension}
\label{section3}

We consider the (negative) bulk cosmological constant and brane tensions and
matters on the brane. In addition, we also consider non-zero $\hat p_5$
component in addition to the bulk cosmological constant.
\begin{equation}
\label{EM}
\hat\rho=\Lambda_b,\quad \hat p=-\Lambda_b,\quad \hat
p_5=-\Lambda_b+p_5(\tau,y).
\end{equation}

Now we assume that the extra dimension is stabilized by some mechanism,
and follow the evolution of the scale factor on the brane
after the extra dimension is settled down at a stable point,
in the sense that $\dot b=0$.
Then we can always make $b$ a constant,
which measures the physical size of the extra dimension.
For the bulk energy momentum (\ref{EM}), the equation (\ref{EMC1}) is
automatically satisfied and the equation (\ref{EMC2}) restricts the form of
$p_5(\tau,y)$ to be
\begin{equation}
p_5(\tau,y) = \frac{\tilde p_5(\tau)}{n(\tau,y)a^3(\tau,y)}.
\end{equation}
For $\dot b=0$, the equation (\ref{05}) implies that $\dot a/n$ is
$y$ independent, that is
\begin{equation}
\label{lambda}
\frac{\dot a(\tau,y)}{n(\tau,y)} \equiv \lambda(\tau),
\end{equation}
and (\ref{C}) becomes
\begin{equation}
\label{C1}
\dot C %= \frac{(a^4)\dot{\vphantom a}}{6M^3}p_5
= \frac{2\lambda(\tau)}{3M^3}\tilde p_5(\tau).
\end{equation}
Now the equation (\ref{F}) with $\hat\rho=\Lambda_b$ can be written as
\begin{equation}
\left(\frac{a'}{b}\right)^2 = 
\frac{-\Lambda_b}{6M^3}a^2 + \left(\frac{\dot a}{n}\right)^2 + K
+\frac{C}{a^2}.
\end{equation}
Since $\dot a/n=\lambda(\tau)$ is $y$-independent for $\dot b=0$,
we can perform $y$-integration again.
For the negative bulk cosmological constant, $\Lambda_b<0$, we obtain
\begin{equation}
\label{a}
a^2(\tau,y) = \frac{D(\tau)}{2k}e^{2kby}
+ \frac{1}{2kD(\tau)}\left[\left(\frac{\lambda^2(\tau)+K}{2k}\right)^2
-C(\tau)\right]e^{-2kby} - \frac{\lambda^2(\tau)+K}{2k^2},
\end{equation}
where $k^2=-\Lambda_b/6M^3$ and $D(\tau)$ is the second integration constant.
We may write $D(\tau)$ in terms of $a_0(\tau)\equiv a(\tau,y=0)$
and $\dot a_0(\tau)\equiv\dot a(\tau,y=0)$
if we fix the gauge by the condition $n(\tau,y=0)=1$.
Then the Eq.~(\ref{lambda}) gives $\lambda(\tau)=\dot a_0(\tau)$,
and the consistency requires $a^2$ to be written as
\begin{eqnarray}
a^2(\tau,y) &=& a_0^2\left[
\left(1+\frac{\dot a_0^2+K}{2k^2a_0^2}\right)\cosh(2kby)
-\frac{\dot a_0^2+K}{2k^2a_0^2}
\right.\nonumber\\ \label{afinal} && \hspace{10mm} \left.
\pm \left(1+\frac{1}{k^2}
\left[\frac{\dot a_0^2+K}{a_0^2}+\frac{C}{a_0^4}\right]
\right)^{1/2}\sinh(2kby) \right].
\end{eqnarray}
Hence we get the whole bulk solution,
once $a_0(\tau)$ and $C(\tau)$ are known.
$C(\tau)$ is determined up to a constant
by the bulk momentum $p_5$ according to the Eq.~(\ref{C1}).
$a_0(\tau)$ can be fixed by a brane boundary through (\ref{BC}).

Suppose that a brane is placed at $y=0$, and assume the $Z_2$ symmetry,
$y\sim-y$, meaning the brane forms a boundary of five-dimensional spacetime.
Then the boundary condition (\ref{BC}) applied to (\ref{afinal}) gives
a evolution equation for $a_0(\tau)$
\begin{equation}
\label{Friedmann}
\left(\frac{\dot a_0}{a_0}\right)^2+\frac{K}{a_0^2}
= -k^2 + \left(\frac{\rho_0}{6M^3}\right)^2 - \frac{C}{a_0^4}.
\end{equation}
This is a Friedmann equation for the scale factor on the brane.
This equation looks different from the usual four dimensional Friedmann
equation, in that the right hand side has a term which is not proportional
to $\rho_0$ but to $\rho_0^2$ and an additional $C$ term which reflects
the influence of bulk matter through the Eq.~(\ref{C1}).
In the absence of bulk matter, that is, $p_5=0$ in our framework,
$C$ is a constant fixed by the initial condition.
The $C$-term gives a radiation-like
(in the sense that it is proportional to $a_0^{-4}$)
contribution to the energy density, and its size can limited by
nucleosynthesis \cite{BDEL}.
The behavior $H^2\propto\rho_0^2$ is much more problematic,
if we assume the brane is our universe,
since it results in unconventional evolution of the scale factor
and spoils good predictions of nucleosynthesis.
This is a cosmological difficulty associated with the simple brane world
scenario \cite{BDL}.

The behavior $H^2\propto\rho_0$ can be recovered by introducing
the negative bulk cosmological constant together with the brane tension.
For the brane with both tension and matter,
replacing $\rho_0$ with $\Lambda_0+\rho_{0M}$ in the Eq.~(\ref{Friedmann}),
we obtain
\begin{equation}
\label{Friedmann1}
\left(\frac{\dot a_0}{a_0}\right)^2+\frac{K}{a_0^2}
= (k_0^2-k^2) %+ \left(\frac{\Lambda_0}{6M^3}\right)^2
+ \frac{\Lambda_0}{18M^6}\rho_{0M}
+ \frac{1}{36M^6}\rho_{0M}^2
- \frac{C}{a_0^4},
\end{equation}
where $k_0=\Lambda_0/6M^3$.
Identifying the four dimensional Planck mass $M_P^2=6M^6/\Lambda_0$
and taking $C=0$,
we recover the usual four dimensional Friedmann equation
for $\rho_{0M}\ll\Lambda_0$.
Adjusting the bulk cosmological constant and the brane tension to cancel each
other, that is $k^2=k_0^2$, corresponds to a fine tuning which makes the
four-dimensional cosmological constant vanish.
This gives a viable cosmology for the positive tension brane attached to
the infinite size extra dimension \cite{RS2,BDEL}.
For the finite size extra dimension, we must consider the effect of
stabilization mechanism for the finite size,
which inevitably introduces bulk matter in the story.
On the other hand, if we have a negative tension brane as in the original
Randall-Sundrum model, this leads to anti-gravity on that brane
since a opposite sign appears in front of $\rho_{0M}$.
Unfortunately, this negative tension brane was
considered to be our universe in the RS model to solve the gauge hierarchy
problem. We will be back to this negative tension brane problem
in the next section.

\section{Stabilization of the extra dimension by balancing bulk matter
and cosmological solutions}
\label{section4}

A quite interesting observation was made in Ref.~\cite{KKOP1},
that the usual four-dimensional Friedman equation is recovered by
the bulk matter which is correlated with the brane matter.
If we have bulk matter, $\hat p_5$ in this paper, of the form
\begin{equation}
\hat p_5 = -\frac{a_0^3}{2na^3}
\left[
    \frac{M^3}{M_P^3}(\rho_0-3p_0)
   +\frac{1}{6M^3}\rho_0(\rho_0+3p_0)
\right],
\end{equation}
it gives rise to a $C$-term
\begin{equation}
C = \left[
    -\frac{\rho_0}{3M_P^2} + \frac{\rho_0^2}{36M^6}
\right] + C_0,
\end{equation}
so that the usual Friedmann equation can be achieved
up to a constant $C_0$ term.
In the subsequent paper by the same authors \cite{KKOP2},
it was argued that this bulk matter distribution arises
from the back reaction to the brane matter
in the stabilization mechanism.

The same idea can be used to resolve the cosmological difficulty
when we have a negative tension brane.
We can think of non-trivial $\hat p_5$ correlated with brane energy densities
in such a way that it gives rises to
\begin{equation}
\label{C2}
\frac{C(\tau)}{a_0^4} = -\frac{\Lambda_0}{9M^6}\rho_{0M} +{\cal O}(\rho_{0M}^2).
\end{equation}
However, this possibility should be considered
together with the stabilization mechanism.

Let us consider the evolution of scale factor in the Randall-Sundrum model.
We put the second brane at $y=\frac12$, and
consider the five-dimensional bulk surrounded by two branes.
Now the boundary condition (\ref{BC}) at $y=\frac12$ imposes a constraint
\begin{equation}
\label{constraint}
\bar\rho_{\frac12} = \frac
{\sinh(kb)+\frac12(\bar\rho_0^2-\bar C-1)\sinh(kb)-\bar\rho_0\cosh(kb)}
{\cosh(kb)+\frac12(\bar\rho_0^2-\bar C-1)(\cosh(kb)-1)-\bar\rho_0\sinh(kb)},
\end{equation}
where we used a notation
$\bar\rho_i=\rho_i/6M^3k$ ($i=0,\frac12$) and $\bar C=C/k^2a_0^4$.
Without bulk matter, i.e., $C={\rm constant}$, it gives a constraint between
energy densities on two branes. This is necessary to keep the extra dimension
static without assuming any stabilization mechanism.
Let us examine this constraint in detail in the RS setup,
where two boundary branes have tensions $\Lambda_i$ ($i=0,\frac12$)
of the same size with opposite signs
and adjusted to satisfy $\Lambda_i^2/6M^3=\Lambda_b$.
We put the negative tension brane at $y=0$,
and replace $\bar\rho_0$ and
$\bar\rho_{\frac12}$ with $-1+\bar\rho_{0M}$ and $+1+\bar\rho_{\frac12M}$
respectively in (\ref{constraint}), to get the constraint in the RS model
\begin{equation}
\label{constraint2}
\bar\rho_{\frac12M} = \frac
{\left(-\bar\rho_{0M}+\frac12\bar\rho_{0M}^2-\frac12\bar C\right)
\left(1-e^{-kb}\right)-\bar\rho_{0M}e^{-kb}}
{e^{kb}+\left(-\bar\rho_{0M}+\frac12\bar\rho_{0M}^2-\frac12\bar C\right)
\left(\cosh(kb)-1\right)
-\bar\rho_{0M}\sinh(kb)}.
\end{equation}
For $\bar\rho_{0M},\bar\rho_{\frac12M}\ll1$, and $e^{kb}\gg1$
which is necessary to solve the gauge hierarchy problem in this framework,
it becomes
\begin{equation}
\label{constraint3}
\bar\rho_{\frac12M}\approx e^{-kb}\left(-\bar\rho_{0M}-\frac12\bar C\right).
\end{equation}
For $C=0$, it is required that
$\rho_{\frac12M}$ and $\rho_{0M}$ have the opposite signs, and furthermore
$\rho_{\frac12M}$ is correlated with $\rho_{0M}$ by the same exponential factor
which is used to solve the gauge hierarchy problem.
This parallels the situation in the inflationary solution that
the more fine balance between the bulk cosmological constant and
the brane tension by the same factor is necessary in the positive tension
brane to keep the extra dimension static, when the relation
$\Lambda_i^2/6M^3=\Lambda_b$ does not hold exactly \cite{KK}.
Csaki et al.~\cite{CGRT} pointed out that this odd constraint is due to
requiring the static extra dimension without proper inclusion of
stabilization mechanism.

Here we attempt to connect the bulk energy momentum $\hat T^5_5$ to the
stabilization mechanism,
as a possible and simple model as to how the stabilization
mechanism works in brane models.
We suppose that there is an (unspecified) stabilization mechanism,
and it gives rise to bulk energy-momentum tensor dynamically,
specifically $\hat T^5_5$,
as a back reaction to the brane energy densities which, if left alone,
would destabilize the extra dimension.
In general, it is expected that the stabilization mechanism induces
$\hat T^i_i$ as well as $\hat T^5_5$ \cite{GW2},
but we focus on the role of $\hat T^5_5$ in this paper.
This means in our framework that the stabilization mechanism gives rise to
$C$ term satisfying the constraint (\ref{constraint})
for any given $\rho_0$ and $\rho_{\frac12}$
which is necessary to keep the extra dimension static.
Solving it for $C$, we get
\begin{equation}
\label{CC}
\bar C = (\bar\rho_0^2-1) - 2\frac
{(-1-\bar\rho_0\bar\rho_{\frac12})\sinh(kb)+(\bar\rho_0+\bar\rho_{\frac12})\cosh(kb)}
{\sinh(kb)-\bar\rho_{\frac12}(\cosh(kb)-1)}.
\end{equation}
There can be a correction to this $C$ term, because $b$ may not be strictly
static but be shifted somewhat as $\rho_i$ changes in time.
We assume that the amount of this shift is very small and the above $C$
and the corresponding $\hat T^5_5$ which induces it
gives a good effective description of the dynamics of stabilization mechanism.

Inserting this into the Eq.~(\ref{Friedmann}),
we obtain a very unusual Friedmann equation for the scale factor
of the brane located at $y=0$
\begin{equation}
\label{Friedmann2}
\left(\frac{\dot a_0}{a_0}\right)^2+\frac{K}{a_0^2} = 2k^2\frac
{(-1-\bar\rho_0\bar\rho_{\frac12})\sinh(kb)+(\bar\rho_0+\bar\rho_{\frac12})\cosh(kb)}
{\sinh(kb)-\bar\rho_{\frac12}(\cosh(kb)-1)}.
\end{equation}
Not only $\rho_0$ but also $\rho_{1/2}$ appears in the right hand side
due to the influence of balancing bulk matter.

Let us look at the case $\rho_{\frac12}=0$ first.
This corresponds to compactifying the extra dimension without the second
brane.  This is made possible by the role of $\hat p_5$.
This case was also considered by Kanti, et al.\ recently~\cite{KKOP2}.
The Friedmann equation (\ref{Friedmann2}) becomes
\begin{equation}
\left(\frac{\dot a_0}{a_0}\right)^2+\frac{K}{a_0^2} =
2k^2\left[-1+\bar\rho_0{\rm coth}(kb)\right],
\end{equation}
and the corresponding $\hat p_5$ is given by
\begin{equation}
\hat p_5 = \frac{a_0^3}{na^3}\left[
    6M^3k^2 - \frac12 k{\rm coth}(kb)(\rho_0-3p_0)
    - \frac{\rho_0(\rho_0+3p_0)}{12M^3}
\right].
\end{equation}
If we take out the brane tension from the brane energy density,
the above equations become
\begin{eqnarray}
\label{Friedmann3}
& \displaystyle
\left(\frac{\dot a_0}{a_0}\right)^2+\frac{K}{a_0^2} =
2k^2\left[\left(\frac{k_0}{k}{\rm coth}(kb)-1\right) +
{\rm coth}(kb)\bar\rho_{0M}\right]
& \\ & \displaystyle
\hat p_5 = \frac{a_0^3}{na^3}\left[
    6M^3\left(k^2+k_0^2-2kk_0{\rm coth(kb)}\right)
\vphantom{-\frac{\rho_0(\rho_0+3p_0)}{12M^3}}
\hspace{30mm}\right.&\nonumber\\&\displaystyle\left.\hspace{30mm}
   -\frac12\left(k{\rm coth}(kb)-k_0\right)(\rho_{0M}-3p_{0M})
   -\frac{\rho_{0M}(\rho_{0M}+3p_{0M})}{12M^3}
\right],
&
\end{eqnarray}
where $k_0=\Lambda_0/6M^3$.
The Eq.~(\ref{Friedmann3}) gives the usual Friedmann equation
for the positive tension brane,
with the identification $M_p^2=(M^3/k)\tanh(kb)$ and
$\Lambda_{\rm eff}=\Lambda_0-(6M^3\Lambda_b)^{\frac12}\tanh(kb)$.
This model does not address the gauge hierarchy problem,
but it has intriguing implication for the cosmological constant.
Suppose that we have the relation $k=k_0$ in some way.
Then the size of the cosmological constant has an exponential dependence
on the size of extra dimension.
It means we can obtain a very tiny cosmological constant
from a moderate size of extra dimension.
For example, $\Omega_\Lambda\sim1$ can be obtained for $kb\approx140$.
The situation is very similar to that of the original RS model which converts
the gauge hierarchy problem to a dynamical problem which determines
the size of extra dimension.
This model does the same thing for the cosmological constant problem.
Here we are not to urge that this model is on the way to a genuine solution
to the cosmological constant problem, since such a small difference can
always overwhelmed by other perturbations.
But it's a good thing to have such a model.

Now we turn to the two brane case.
Splitting the brane energy density into
the brane tension and the matter energy density,
The Friedmann equation (\ref{Friedmann2}) can be rewritten as
\begin{eqnarray}
\left(\frac{\dot a_0}{a_0}\right)^2+\frac{K}{a_0^2} &=&
\frac{2k^2}{e^{kb}-1
   -\left(\frac{k_{\frac12}-k}{k}+\bar\rho_{\frac12M}\right)e^{kb}
    \left(\cosh(kb)-1\right)}
\nonumber\\ && \hspace{-20mm} \times
\left[
    \left(\frac{k+k_0}{k}\right)\left(\frac{k_{\frac12}+k}{2k}\right)
   +\left(\frac{k_{\frac12}-k}{k}\right)\left(\frac{k-k_0}{2k}\right)e^{2kb}
   +\bar\rho_{0M}\left(1-\frac{k_{\frac12}-k}{k}e^{kb}\sinh(kb)\right)
\right. \nonumber\\ && \hspace{-10mm}\left.
   +\bar\rho_{\frac12M}e^{2kb}\left(1-\frac{k+k_0}{k}e^{-kb}\sinh(kb)\right)
   +\bar\rho_{0M}\bar\rho_{\frac12M}e^{kb}\sinh(kb)
\right]
\end{eqnarray}
where $k_i=\Lambda_i/6M^3$ and $\Lambda_i$, ($i=0,\frac12$) are brane tensions.
Note that, because we set $n=1$ at the negative tension brane,
the mass parameters in the action are of order of the electroweak scale.
The Planck mass is given by $M_P\approx(M^3/k)e^{kb}$,
which differs by a $e^{kb}$ factor from that in the RS paper,
and the physical brane tension and energy density of the positive tension
brane are scaled accordingly.
If this model has anything to do with our universe up to the electroweak
scale, and to solve the gauge hierarchy problem,
the orders of magnitude of parameters have to satisfy
\begin{eqnarray}
&\displaystyle
\left(\frac{k_{\frac12}-k}{k}\right)e^{2kb}
\sim \left(\frac{k+k_0}{k}\right) \lesssim
\left(\frac{\rho_c}{M_P}\right)^4 \sim 10^{-120},
& \nonumber\\ \label{estimate} &\displaystyle
\bar\rho_{0M},\bar\rho_{\frac12M}e^{2kb} \lesssim
\left(\frac{M_W}{M_P}\right)^4 \sim 10^{-64},
&\\&\displaystyle
e^{-kb} \sim \frac{M_W^2}{M_P^2} \sim 10^{-32}, 
& \nonumber
\end{eqnarray}
where $\rho_c$ is the critical density and $M_W$ is the electroweak scale.
When the strict Randall-Sundrum condition $k=-k_0=k_{\frac12}$ is imposed, the
cosmological constant term vanishes and the above equation is simplified to
\begin{equation}
\label{Friedmann4}
\left(\frac{\dot a_0}{a_0}\right)^2+\frac{K}{a_0^2} =
\frac{\rho_{0M}+\rho_{\frac12M}e^{2kb}
-\rho_{0M}\bar\rho_{\frac12M}e^{kb}\sinh(kb)}
{3(M^3/k)\left[e^{kb}-1-\bar\rho_{\frac12M}e^{kb}(\cosh(kb)-1)\right]}.
\end{equation}
If we consider the situation where
matter resides only on the negative tension brane,
that is, $\rho_{\frac12M}=0$ (or $\rho_{\frac12M}e^{2kb}\ll\rho_{0M}$),
we obtain nothing but the four-dimensional Friedmann equation.
The stabilization mechanism,
through the balancing $\hat T^5_5$ component in this simplified model,
resolves the wrong sign of negative tension brane Friedman equation
and reproduce ordinary FRW cosmology for the brane.
When there is non-vanishing matter density on the positive tension brane,
it acts as dark matter for the negative tension brane,
as is generically expected.
Finally, for the cosmological constant, we noted above that the cosmological
constant vanishes if the Randall-Sundrum condition exactly holds.
If there is really a small cosmological constant in our universe,
the balance between the bulk cosmological constant and the brane tension
must be adjusted not only on the negative tension brane but also
on the positive tension brane, in the presence of the stabilization mechanism.

\section{Conclusion}
\label{section5}

In conclusion,
we provided an exact five-dimensional bulk solution
which corresponds to a cosmological solution of RS models,
when the bulk energy momentum $\hat T^5_5$ is incorporated.
We showed that the existence of bulk energy-momentum $\hat T^5_5$
alters the evolution of brane scale factor in an interesting way,
and can be used to stabilize the extra dimension in RS models.
We exploited the idea that $\hat T^5_5$ is adjusted dynamically
in such a way to stabilized the extra dimension.
This is a simple and possible way how the stabilization mechanism works
in the RS models.
Its impact on the evolution of brane scale factor is quite intriguing.
We can construct a single brane RS model with a compact extra dimension.
This model has a nice feature that the effective cosmological constant
on the brane depends exponentially on the size of extra dimension,
thereby it provides a RS-type solution to the cosmological constant problem.
For the two brane RS model, the balancing $\hat T^5_5$ makes it possible
to reproduce ordinary FRW cosmology for the RS model, resolving the wrong sign
in the Friedmann equation for the negative tension brane.
Finally, we wish to mention possible loopholes
in the models considered in this paper.
First, we took only $\hat T^5_5$ into consideration in this paper,
but it is plausible that the stabilization mechanism induces $\hat T^i_i$
as well as $\hat T^5_5$. Inclusion of induced $\hat T^i_i$ may alter the
conclusions presented here. Second, we required a strictly static extra
dimension $\dot b=0$ during the evolution of brane,
but this is not obligatory for phenomenology,
nor strictly achieved by the stabilization mechanism.

\acknowledgements

We thank K. Choi, H. D. Kim, and D. Lyth for useful discussions and comments.
H. B. Kim is supported by PPARC grant GR/L40649.

\select{}{\end{multicols}}
\end{document}